\documentclass[conference]{IEEEtran}

\usepackage{float}
\usepackage{enumitem}
\usepackage{cite}
\usepackage{graphicx,amsmath,amssymb,url,multirow}
\usepackage{lipsum}
\ifCLASSOPTIONcompsoc
  \usepackage[caption=false,font=normalsize,labelfont=sf,textfont=sf]{subfig}
\else
  \usepackage[caption=false,font=footnotesize]{subfig}
\fi

\makeatletter

\begin{document}

\title{Joint Estimation of System Inertia and Load Relief}

\author{\IEEEauthorblockN{Julius Susanto \textit{Senior Member, IEEE}, Alireza Fereidouni \textit{Member, IEEE} and Dean Sharafi \textit{Senior Member, IEEE}}
\IEEEauthorblockN{System Management, Australian Energy Market Operator, Perth, WA 6000, Australia}}

\maketitle

\begin{abstract}
Modern power systems with high share of renewable generation are at the risk of rapid changes in frequency resulting from contingencies. The importance of an accurate assessment of system inertia and load relief, as well as frequency changes in the critical timescales is more pronounced in these power systems. This knowledge serves as an insight which will guide the solutions required for enhanced security and resilience. A novel procedure for simultaneously assessing system inertia and load relief using a system frequency response (SFR) model to fit measured disturbance data is presented. Results of applying the proposed approach on generator contingencies in the South West Interconnected System (SWIS) in Western Australia demonstrate the validity of the method and indicate that it can overcome some of the limitations observed in conventional inertia estimation methods based on the linearised swing equation, such as the sliding window and polynomial fit methods.
\end{abstract}

\vspace{1em}


\section{Introduction}
Many power systems across the world are experiencing a gradual decline in synchronous inertia levels as synchronous generators are increasingly being displaced by generation interfaced to the grid via power-electronic converters \cite{hartmann_2019}. As inertia levels decrease, the average frequency in the power system is subject to more severe and rapid fluctuations in response to active power disturbances, e.g. from generation or load contingencies. It is therefore becoming more crucial that estimates of system inertia are reliable and accurate.


Most system inertia estimation techniques, whether offline estimates such as \cite{Inoue_1997}, \cite{ENTSOE_2018}, \cite{chassin_2005} and \cite{ashton_2015} or online estimates such as \cite{Wall_2014}, \cite{Cai_2019} and \cite{Wilson_2019}, rely on a linearised representation of the swing equation evaluated at the onset of a disturbance ($t=0$):

\begin{equation}
    \label{eqn:inertia_estimate}
    \overline{KE} = -\frac{f_{n}}{2} \times \frac{P_{cont}}{ \frac{\overline{d \Delta f}}{dt}\bigg|_{t=0}}
\end{equation}

\noindent where $\overline{KE}$ is the system inertia estimate (MW.s), $f_n$ is the nominal frequency (e.g. $50$ Hz), $P_{cont}$ is the disturbance size (MW) and $\frac{d\overline{\Delta f}}{dt} \big|_{t=0}$ is an estimate of the rate of change of frequency (RoCoF) evaluated at the onset of the disturbance (Hz/s). 

For this linearised representation to be valid, it is assumed that i) there are no frequency deviations at the onset of the disturbance, i.e. $\Delta f(t=0) = 0$, ii) primary frequency response has yet to begin acting, and ii) load relief is negligible in the early stages after a disturbance. To satisfy these assumptions, inertia estimates of the form described in (\ref{eqn:inertia_estimate}) are often performed using historical generator contingency events, filtered to only select for single instantaneous disturbances (e.g. protection trip) as opposed to ramp downs (e.g. due to a generating unit boiler malfunction / failure) or contingencies involving multiple generating units. With such sudden single unit trips, the contingency size $P_{cont}$ can normally be estimated quite accurately \cite{ashton_2015}. Alternatively, known active power perturbations by normal operational events such as the reversal of power in an HVDC interconnector \cite{best_2021} or an intentionally injected signal \cite{hosaka_2019} can also be used.

\begin{figure}[htp]
\centering
\includegraphics[width=\linewidth]{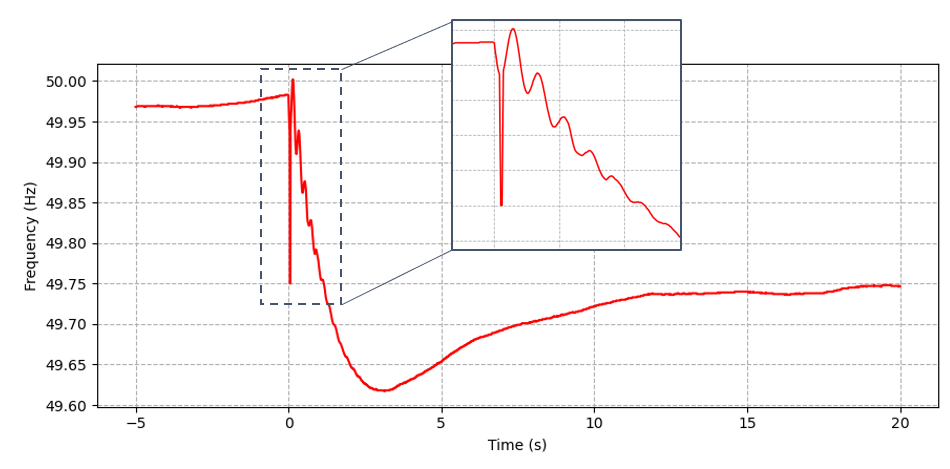}
\caption{Transients and oscillations in frequency measurements}
\label{fig:freq_meas_error}
\end{figure}

However, the RoCoF is arguably the most sensitive parameter in (\ref{eqn:inertia_estimate}). Ideally, the RoCoF would be calculated based on the sample-by-sample time derivative of a high-resolution frequency trace at the onset of the disturbance. However, high-resolution frequency measurements from Phasor Measurement Units (PMUs) or fault recorders contain white noise introduced by transducer, quantisation and signal processing errors \cite{Brown_2016}. Moreover, as depicted in Fig. \ref{fig:freq_meas_error}, the frequency trace can contain transient \cite{ashton_2015} and/or oscillatory components \cite{Inoue_1997}. Therefore, calculating the time derivative using consecutive frequency samples would likely lead to significant errors \cite{ENTSOE_2018}. 

To mitigate these issues, moving average windows or low-pass smoothing filters have been proposed in the literature, as well as in practical applications. For example, a sliding window has been adopted as a standard in Europe (e.g. \cite{ENTSOE_2018} and \cite{ashton_2015}), where the RoCoF estimate is calculated by applying a 500-ms sliding window over the frequency trace and then finding the maximum value of the sample-by-sample time derivative (since the maximum RoCoF theoretically occurs at the onset of the contingency). Alternative techniques include fitting a polynomial function to the raw frequency measurements \cite{Inoue_1997} or simply applying a 0.5 Hz low-pass filter to the frequency trace \cite{chassin_2005}. The common thread is that the raw frequency measurements are smoothed out in some way to prevent spurious RoCoF measurements.

It was observed in \cite{susanto_2019} that system inertia estimates based on this approach are only accurate for events with relatively high RoCoFs, in which the assumption that the initial frequency decline is retarded by system inertia alone remains valid. For smaller disturbances or high inertia conditions where the RoCoF is low, the influence of primary frequency response becomes more prominent and may invalidate this assumption.  Furthermore, the choice of sliding window length (or cut-off frequency for the smoothing filter) will also affect the quality of the estimate, with higher levels of smoothing leading to larger estimation errors. It was also shown that inertia can be materially overestimated when wind farms provide virtual inertia emulation or fast frequency response \cite{fernandez_2019}.

Another factor that may confound system inertia estimates is the presence of load relief (or damping), which is the natural sensitivity of loads in a network to changes in system frequency. As frequency declines after a generation contingency, so does power consumption from frequency-sensitive loads such as induction motors \cite{masood_2018}. Load relief acts to correct the active power imbalance after a contingency and thus helps to arrest the decline in frequency. As a result, system inertia estimates based on the linearised swing equation after a disturbance may mistake the effects of load relief as additional inertia.

In the literature, the system-level load relief factor (LRF), which is also referred to as the load damping coefficient or load-frequency characteristic, is typically estimated separately from system inertia. Earlier studies attempted to directly measure load relief by conducting live tests such as the intentional tripping of generators and transmission lines, e.g. in Great Britain \cite{davies_1959}, Norway \cite{berg_1972} and Ireland \cite{osullivan_1996}. A less invasive approach is to estimate the system-level LRF after a contingency event using measurements of total system load from the Energy Management System (EMS). However, EMS sampling resolution may not be high enough to accurately assess system load changes after a contingency \cite{pearmine_2006}. To mitigate EMS resolution issues, high-speed fault recorder data can be used to estimate the system load change after a contingency \cite{susanto_2019}. Finally, estimating LRF has also been investigated using a bottom-up approach, e.g. using high-speed fault recorder data at substation level \cite{masood_2018} or at the level of individual consumer devices \cite{Omara_2012}. However, it is remains unclear how to aggregate the individual LRF estimates into a meaningful system-level LRF. 

This paper takes a different approach to system inertia and load relief estimation. Rather than using the linearised swing equation, the system inertia and load relief factor are treated as unknown parameters in a low-order system frequency response (SFR) model and the model is then fitted to post-disturbance frequency and generator output measurements. This approach exploits the SFR model's ability to accurately predict the system frequency trajectory after a disturbance, and thus use more measurement data as inputs to the estimation model.


The main contribution of this paper is the introduction of a novel model fitting approach to inertia estimation that uses a system frequency response model to simultaneously estimate both system inertia and system-level load relief. The proposed method is shown to be credible when compared with other inertia estimation methods, while requiring less discretion in the choice of parameters, e.g. sliding window length or polynomial order.

The remainder of this paper is organised as follows: In Section II, the system inertia estimation method based on model-fitting is proposed, with the added benefit that the method concurrently estimates the system-level load relief factor. This is followed in Section III by several computer simulation case studies comparing the performance of the proposed approach with other methods. In Section IV, case studies are performed using data from a real power system to illustrate some of the advantages of the proposed method over other approaches. A discussion of the results in this paper is then presented in Section V, and finally concluding remarks and avenues for future work are described in Section VI.

\section{Proposed Methodology}

Rather than using the linearised swing equation in (\ref{eqn:inertia_estimate}) to estimate system inertia, we propose to jointly estimate system inertia and load relief / damping by fitting a low-order SFR model to measured frequency data.

At a high level, the proposed model-fitting estimation method has two steps as follows:
\begin{enumerate}
    \item Compile generator active power measurement traces to form an aggregate system primary frequency response (PFR) trace and contingency trace
    \item Estimate system inertia and load relief factor by fitting an SFR model to the measured frequency trace
\end{enumerate}

\subsection{Step 1: Measured system PFR and contingency traces}
The measured system PFR trace is formed by aggregating the active power measurement traces from individual generators and interruptible loads. In order for the measured system PFR to be accurate (i.e. representative of the actual system PFR), the following conditions should be met:
\begin{itemize}
    \item Active power measurements are high resolution and time-synchronised, e.g. from fault recorders or PMUs with time synchronisation via GPS or IEEE Std 1588 Precision Time Protocol \cite{IEEE_1588}. 
    \item There are measurements at each major generating facility and interruptible load in the system. Where this is not practical, there should be at least measurements at each of the main PFR providers.
\end{itemize}

Note that with this approach, the influence of any PFR not explicitly included in the measured system PFR trace will be considered as load relief. This could include the influence of behind-the-meter embedded generation or small distribution-connected generation that may not be visible by fault recorders or PMUs. 

Similarly, the contingency trace is formed by extracting the active power measurement trace of the contingency (e.g. power flow from a generator or radial network element).

\subsection{Step 2: Estimation of system inertia and LRF via model fitting}
The linear ordinary differential equation (ODE) for a generic low-order SFR model after a contingency can be formulated from the swing equation as follows:

\begin{equation}
    \label{eqn:SMM_ode}
    \frac{d\Delta f(t)}{dt} = \frac{f_n}{2 KE} \left[ p(t) - P_{cont}(t) - D P_{load} \Delta f(t) \right] 
\end{equation}

\noindent where $\Delta f(t)$ is the frequency deviation at time $t$ (Hz), $f_n$ is the nominal frequency (e.g. $50$ Hz), $KE$ is the post-contingent system inertia (in MW.s), $P_{cont}(t)$ is the contingency at time $t$ (MW), $p(t)$ is the aggregate system PFR at time $t$ (MW), $D$ is the load relief factor (\% MW/Hz) and $P_{load}$ is the system load at the onset of the contingency (MW).

Equation (\ref{eqn:SMM_ode}) can be rewritten in discrete form with a fixed time step $\Delta t$ as follows:

\begin{equation}
    \label{eqn:SMM_ode_discrete}
    \Delta f_{k} = \Delta f_{k-1} + \frac{f_n}{2 KE} \left[ p_{k-1} - P_{cont, k-1} - D P_{load} \Delta f_{k-1} \right] \Delta t
\end{equation}

In (\ref{eqn:SMM_ode_discrete}), the measured system PFR trace $p_{k}$ and contingency trace $P_{cont,k}$ were compiled in Step 1 and are known values, as is the system load pre-contingency $P_{load}$. This leaves the system inertia $KE$ and load relief factor $D$ as the unknown parameters to be estimated.

There are many ways to solve this parameter estimation problem, for example using non-linear least squares \cite{burth_1999}, Newton-Raphson algorithms \cite{karrari_2004} or with Kalman filters \cite{ariff_2015}. In this paper, a recursive non-linear least squares approach similar to the method described in \cite{pourbeik_2009} is used to minimise an error function based on the Root Mean Squared Error (RMSE):

\begin{equation}
    \min \left\{ \sqrt{\frac{1}{N} \sum_{k=0}^{N} (\Delta f_{k} - \Delta \hat{f_{k}})^{2} } \right\}
\end{equation}

\noindent where $f$ is the measured frequency, $\hat{f}$ is the frequency calculated from the SFR model and $N$ is the total number of samples.

\section{Computer Simulations}

Computer simulations were performed on the IEEE 9-bus and New England IEEE 39-bus systems to compare the performance of the proposed model fitting approach with two commonly used inertia estimation methods: 
\begin{enumerate}
    \item Sliding window method \cite{ENTSOE_2018}
    \item Inoue polynomial fit method \cite{Inoue_1997}
\end{enumerate}

The time-domain RMS simulations were conducted in DIgSILENT PowerFactory v2020 software and in each case, loads were modelled to be frequency-dependent with a load relief factor of $D = 4 \%$.

\subsection{Case 1: IEEE 9-Bus System}

In Case 1, the IEEE 9-bus system depicts a small islanded system with three generators and 315 MW of load. The default IEEE 9-bus system was modified to include a standard turbine-governor model (TGOV) for each generator. To simulate a frequency disturbance, generator G3 was tripped from 85 MW. 

A summary of the best and worst inertia and load relief estimates for each of the different methods is shown in Table \ref{tab:case1_summary}. It should be noted that only a single estimate is shown for the model fitting approach as there are no adjustable parameters (such as window length or polynomial order).

\begin{table}[htp]
    \centering
    \caption{Case 1: Summary of results}
    \begin{tabular}{cccl}
        \textbf{Method} & \textbf{KE (MW.s)} & \textbf{D (\%)} & \textbf{Remarks} \\
        \hline
        Actual & 1,522 & 4.00 \% &  \\
        Sliding window (best) & 1,524 & - & 60-ms window \\
        Sliding window (worst) & 2,810 & - & 1,000-ms window \\
        Inoue (best) & 1,547 & - & 20\textsuperscript{th} order \\
        Inoue (worst) & 1,730 & - & 14\textsuperscript{th} order \\
        Model fitting & 1,589 & 3.45 \% &  \\
        \hline
    \end{tabular}
    \label{tab:case1_summary}
\end{table}

\begin{figure}[!thp]
    \centering
  \subfloat[Sliding window method (variable window length)\label{fig:case1a}]{%
       \includegraphics[width=\linewidth]{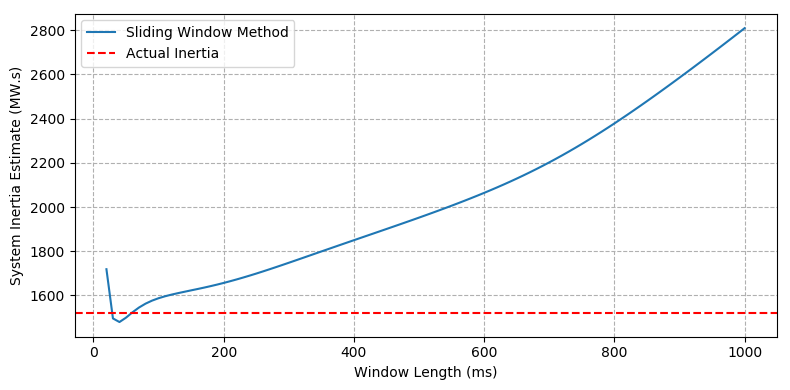}}
    \hfill
  \subfloat[Inoue method (variable polynomial order)\label{fig:case1b}]{%
        \includegraphics[width=\linewidth]{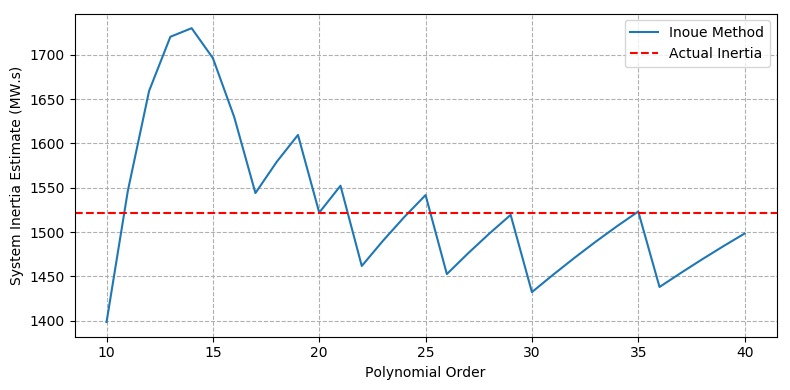}}
    \\
  \subfloat[Model fitting method frequency fit\label{fig:case1c}]{%
        \includegraphics[width=\linewidth]{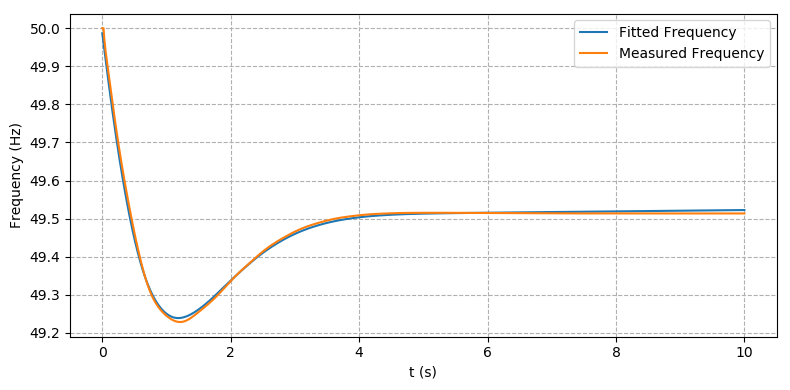}}
    \hfill
  \caption{Case 1: IEEE 9-bus system}
  \label{fig:case1} 
\end{figure}

System inertia estimates using the sliding window and Inoue polynomial fit methods with varying window lengths and polynomial order are shown in Fig. \ref{fig:case1a} and Fig. \ref{fig:case1b} respectively. The figures indicate that the accuracy of the inertia estimate is sensitive to the choice of window length or polynomial order. Fig. \ref{fig:case1c} shows the simulated (measured) frequency trace vs the fitted frequency trace after applying the model fitting method.

\subsection{Case 2: New England IEEE 39-Bus System}

Relative to Case 1, the New England IEEE 39-bus system depicts a larger interconnected power system comprising nine generators and 6,097 MW of load. To simulate a frequency disturbance, generator G10 was tripped from 200 MW. 

\begin{figure}[!thp]
    \centering
  \subfloat[Sliding window method (variable window length)\label{fig:case2a}]{%
       \includegraphics[width=\linewidth]{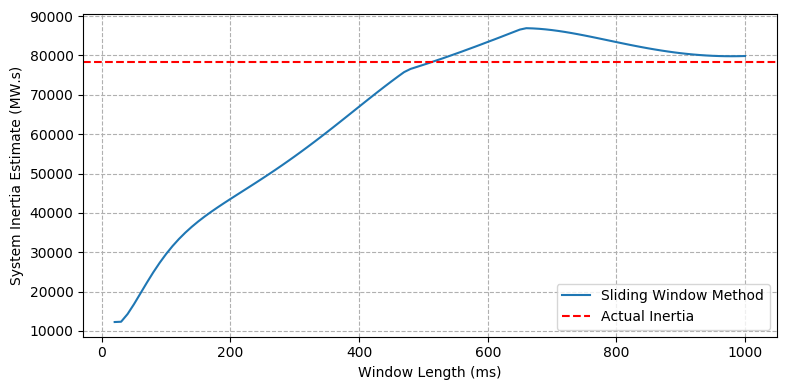}}
    \hfill
  \subfloat[Inoue method (variable polynomial order)\label{fig:case2b}]{%
        \includegraphics[width=\linewidth]{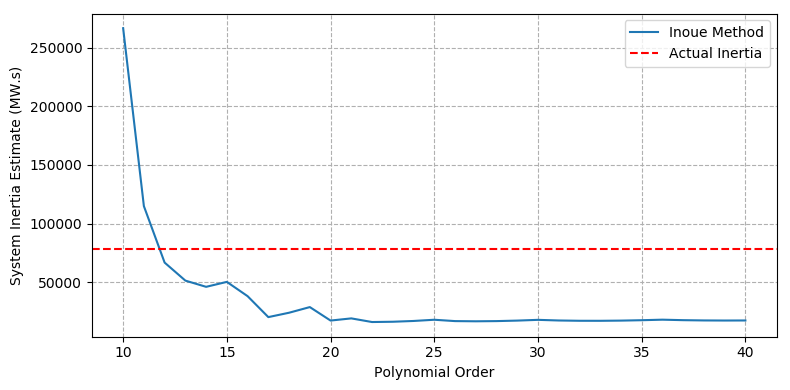}}
    \\
  \subfloat[Model fitting method frequency fit\label{fig:case2c}]{%
        \includegraphics[width=\linewidth]{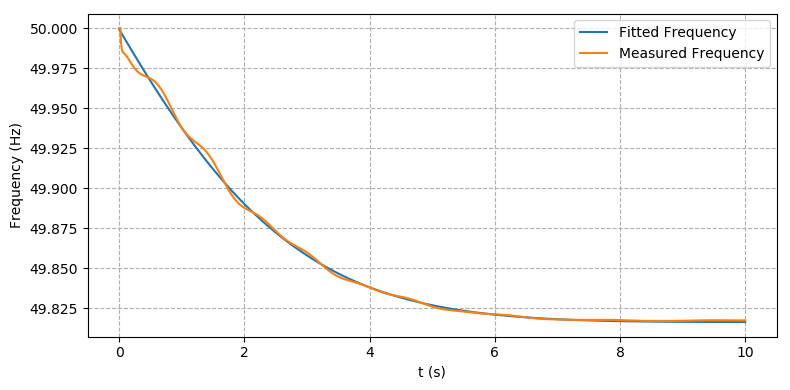}}
    \hfill
  \caption{Case 2: New England IEEE 39-bus system}
  \label{fig:case2} 
\end{figure}

Like in Case 1, a summary of the best and worst inertia and load relief estimates from each of the different methods is shown in Table \ref{tab:case2_summary}. The inertia estimates with varying window length and polynomial order are shown in Fig. \ref{fig:case2a} and Fig. \ref{fig:case2b} respectively, and the measured vs fitted frequency trace after applying the model fitting method is shown in Fig. \ref{fig:case2c}.

\begin{table}[!htp]
    \centering
    \caption{Case 2: Summary of results}
    \begin{tabular}{cccl}
        \textbf{Method} & \textbf{KE (MW.s)} & \textbf{D (\%)} & \textbf{Remarks} \\
        \hline
        Actual & 78,270 & 4.00 \% &  \\
        Sliding window (best) & 78,161 & - & 510-ms window \\
        Sliding window (worst) & 12,270 & - & 20-ms window \\
        Inoue (best) & 66,443 & - & 12\textsuperscript{th} order \\
        Inoue (worst) & 266,772 & - & 10\textsuperscript{th} order \\
        Model fitting & 75,499 & 3.44 \% &  \\
        \hline
    \end{tabular}
    \label{tab:case2_summary}
\end{table}

An interesting feature of the simulation results in Case 2 is the presence of oscillations in the frequency trace as generators in the system swing against each other (see Fig. \ref{fig:case2c}). The oscillations cause the sliding window and polynomial fit methods to overestimate the RoCoF at shorter window lengths (as observed in \cite{ENTSOE_2018}) and high polynomial orders (i.e. because of overfitting). 

\subsection{Case 3: IEEE 9-Bus System with Fast Frequency Response}

In Case 3, the IEEE 9-bus system in Case 1 is augmented with a 30 MW BESS providing FFR. This case is included to examine the effects of FFR on inertia estimates and validate the findings in \cite{fernandez_2019}, i.e. material errors in inertia estimates can arise when the sliding window method is applied to systems with FFR.

\begin{figure}[!thp]
    \centering
  \subfloat[Sliding window method (variable window length)\label{fig:case3a}]{%
       \includegraphics[width=\linewidth]{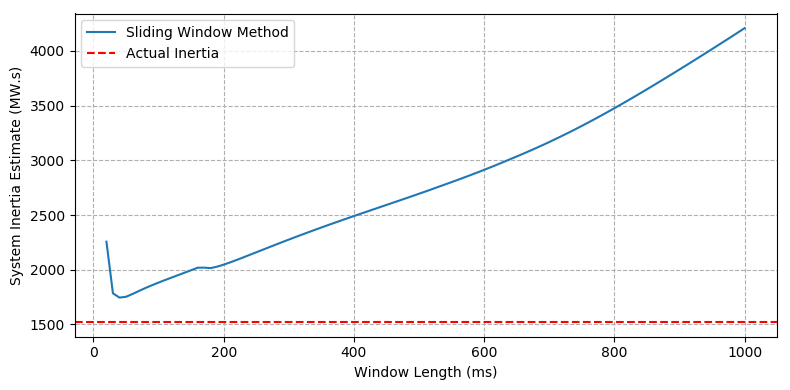}}
    \hfill
  \subfloat[Inoue method (variable polynomial order)\label{fig:case3b}]{%
        \includegraphics[width=\linewidth]{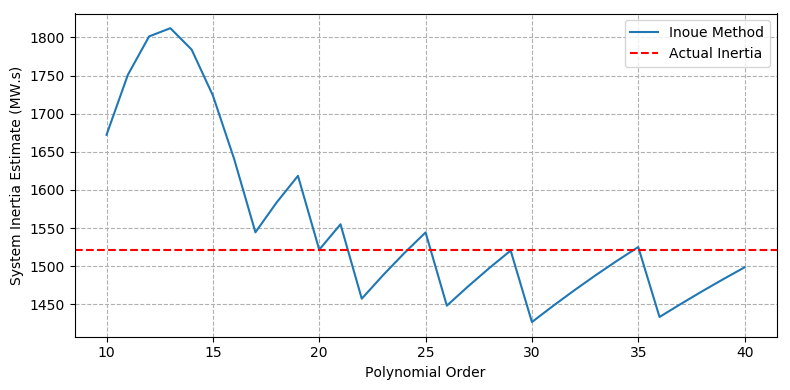}}
    \\
  \subfloat[Model fitting method frequency fit\label{fig:case3c}]{%
        \includegraphics[width=\linewidth]{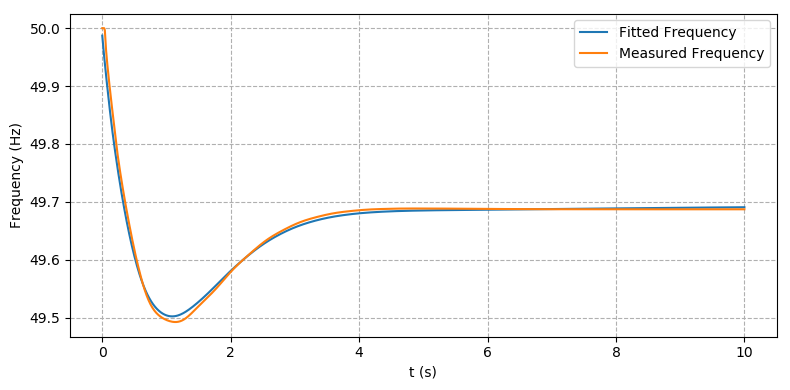}}
    \hfill
  \caption{Case 3: IEEE 9-bus system with FFR}
  \label{fig:case3} 
\end{figure}

As in the previous cases, a summary of the best and worst inertia and load relief estimates from each of the different methods is shown in Table \ref{tab:case3_summary}. The inertia estimates with varying window length and polynomial order are shown in Fig. \ref{fig:case3a} and Fig. \ref{fig:case3b} respectively, and the measured vs fitted frequency trace after applying the model fitting method is shown in Fig. \ref{fig:case3c}.

The results indicate a deterioration in estimation accuracy from the sliding window method from Case 1, particularly with increasing window length. The Inoue polynomial fit method performs better, with a narrower range of estimates, but the accuracy is sensitive to the choice of polynomial order and it may be difficult to ascertain a priori whether the order selected is the appropriate one.

\begin{table}[htp]
    \centering
    \caption{Case 3: Summary of results}
    \begin{tabular}{cccl}
        \textbf{Method} & \textbf{KE (MW.s)} & \textbf{D (\%)} & \textbf{Remarks} \\
        \hline
        Actual & 1,522 & 4.00 \% &  \\
        Sliding window (best) & 1,785 & - & 30-ms window \\
        Sliding window (worst) & 4,209 & - & 1,000-ms window \\
        Inoue (best) & 1,522 & - & 20\textsuperscript{th} order \\
        Inoue (worst) & 1,812 & - & 13\textsuperscript{th} order \\
        Model fitting & 1,662 & 3.76 \% &  \\
        \hline
    \end{tabular}
    \label{tab:case3_summary}
\end{table}

\section{Application to a Real Power System}
\label{sec:applications}
The following case studies are based on actual generator contingency events from the South West Interconnected System (SWIS) in Australia. The SWIS is a medium-sized islanded power system serving approximately 1.2 million customers in the southwest region of Western Australia. The SWIS covers a vast geographic area of 261,000 square kilometres (larger than the land area of the United Kingdom), but with a historical record peak demand of just 4.3 GW. 

\begin{table}[htp]
    \centering
    \caption{Actual contingency case descriptions}
    \begin{tabular}{cl}
         \textbf{Case} & \textbf{Description} \\
        \hline
        Case 4 & Generator prime-mover trip \\
        Case 5 & Network fault causing a generator trip \\
        \hline
    \end{tabular}
    \label{tab:cases}
\end{table}

\subsection{Case 4: Generator turbine trip}

On 17 March 2020 at 18:06pm, an open-cycle gas turbine generator tripped from 161 MW causing frequency to drop to 49.55 Hz (at the nadir). At the onset of the contingency, the system load was 2,441 MW and the total post-contingent generator inertia was 15,009 MW.s (calculated by summing the known generator inertia values for all online machines). 

\begin{figure}[!thp]
    \centering
  \subfloat[Sliding window method (variable window length)\label{fig:case4a}]{%
       \includegraphics[width=\linewidth]{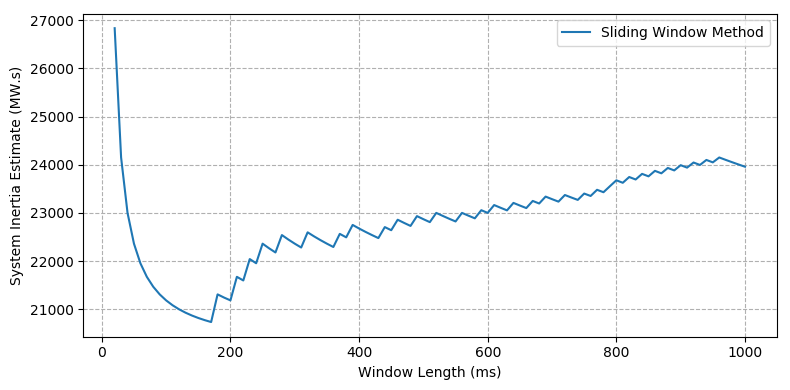}}
    \hfill
  \subfloat[Inoue method (variable polynomial order)\label{fig:case4b}]{%
        \includegraphics[width=\linewidth]{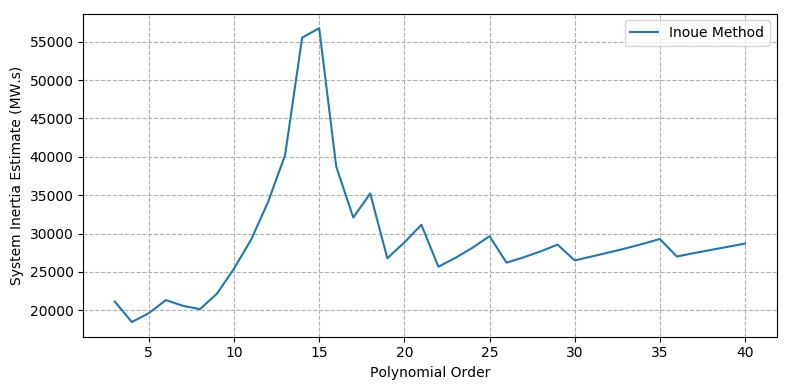}}
    \\
  \subfloat[Model fitting method frequency fit\label{fig:case4c}]{%
        \includegraphics[width=\linewidth]{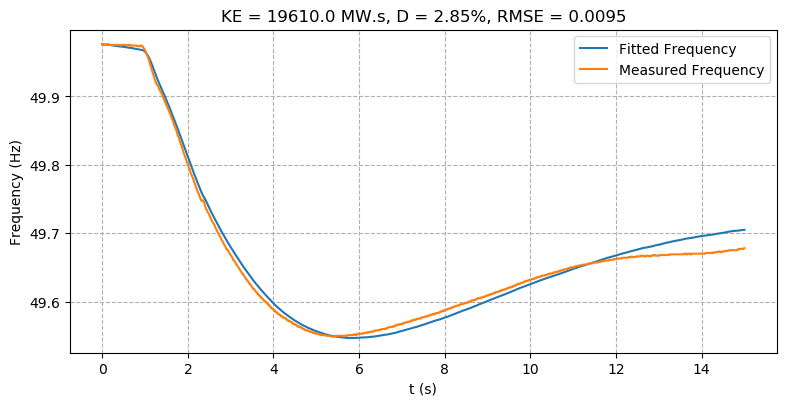}}
    \hfill
  \caption{Case 4: Generator turbine trip}
  \label{fig:case4} 
\end{figure}

The inertia estimates with varying window length and polynomial order are shown in Fig. \ref{fig:case4a} and Fig. \ref{fig:case4b} respectively, and the measured vs fitted frequency trace after applying the model fitting method is shown in Fig. \ref{fig:case4c}.

Unlike in the computer simulations, the actual system inertia is not known. The post-contingent generator inertia is known, but this does not include load inertia so is more representative of the floor for the system inertia estimate. The summary of the results in Table \ref{tab:case4_summary} show the range (minimum and maximum) of estimates from the sliding window and Inoue methods, as well as the single point estimate from the model fitting method.

\begin{table}[htp]
    \centering
    \caption{Case 4: Summary of results}
    \begin{tabular}{cccl}
        \textbf{Method} & \textbf{KE (MW.s)} & \textbf{D (\%)} & \textbf{Remarks} \\
        \hline
        Generator inertia & 15,009 & - &  \\
        Sliding window (min) & 20,735 & - & 170-ms window \\
        Sliding window (max) & 26,833 & - & 20-ms window \\
        Inoue (min) & 18,482 & - & 4\textsuperscript{th} order \\
        Inoue (max) & 56,750 & - & 15\textsuperscript{th} order \\
        Model fitting & 19,610 & 2.85 \% &  \\
        \hline
    \end{tabular}
    \label{tab:case4_summary}
\end{table}

The sliding window method in Fig. \ref{fig:case4a} exhibits a similar shape to that in Cases 1 and 3, where beyond a threshold window length (in this case 170-ms), the inertia estimate continues to rise steadily without settling. In Cases 1 and 3, the actual system inertia was close to the estimate at the threshold window length, which in this case would be 20,735 MW.s.

At high polynomial orders, the Inoue method in Fig. \ref{fig:case4b} appears to settle at around 26,000 MW.s. However, such an estimate would suggest a load inertia of roughly 42\% of system inertia, which is inconsistent with previous inertia estimates in the SWIS, as well as estimates reported in the literature (such as \cite{tavakoli_2012} and \cite{bian_2018}), where load inertia is typically estimated to be 10 - 25\% of system inertia.

The model fitting results in Fig. \ref{fig:case4c} show good alignment between the simulated and measured frequency traces ($RMSE=0.0095$) and the inertia estimate of 19,610 MW.s is on the lower end of the sliding window and and Inoue estimate range. This suggests that the actual system inertia is likely in the 19,000 - 21,000 MW.s range.

\subsection{Case 5: Network fault causing a generator trip}

On 22 October 2020 at 16:25pm, a 330 kV network fault caused the trip of a combined-cycle gas power plant, resulting in an instantaneous loss of 231 MW and frequency to fall to 49.46 Hz (at the nadir). At the onset of the contingency, the system load was 1,991 MW and the total post-contingent generator inertia was 11,612 MW.s. The summary of the results in Table \ref{tab:case5_summary} show the range (minimum and maximum) of estimates from the sliding window and Inoue methods, as well as the single point estimate from the model fitting method.

\begin{table}[htp]
    \centering
    \caption{Case 5: Summary of results}
    \begin{tabular}{cccl}
        \textbf{Method} & \textbf{KE (MW.s)} & \textbf{D (\%)} & \textbf{Remarks} \\
        \hline
        Generator inertia & 11,612 & - &  \\
        Sliding window (min) & 3,609 & - & 40-ms window \\
        Sliding window (max) & 18,392 & - & 1,000-ms window \\
        Inoue (min) & 6,567 & - & 30\textsuperscript{th} order \\
        Inoue (max) & 22,309 & - & 3\textsuperscript{rd} order \\
        Model fitting & 15,534 & 2.01 \% &  \\
        \hline
    \end{tabular}
    \label{tab:case5_summary}
\end{table}

\begin{figure}[!thp]
    \centering
  \subfloat[Sliding window method (variable window length)\label{fig:case5a}]{%
       \includegraphics[width=\linewidth]{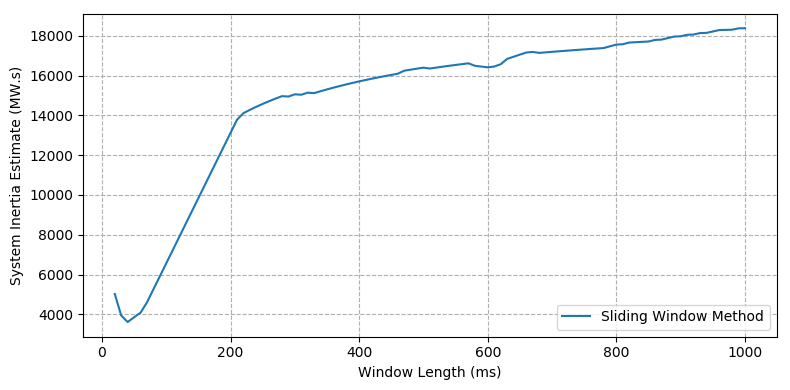}}
    \hfill
  \subfloat[Inoue method (variable polynomial order)\label{fig:case5b}]{%
        \includegraphics[width=\linewidth]{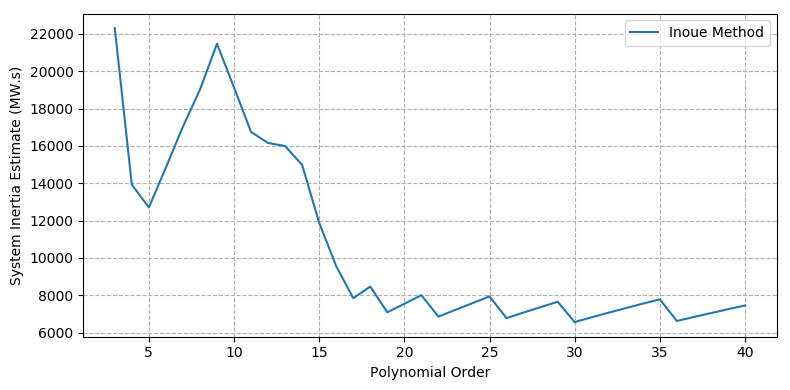}}
    \\
  \subfloat[Model fitting method frequency fit\label{fig:case5c}]{%
        \includegraphics[width=\linewidth]{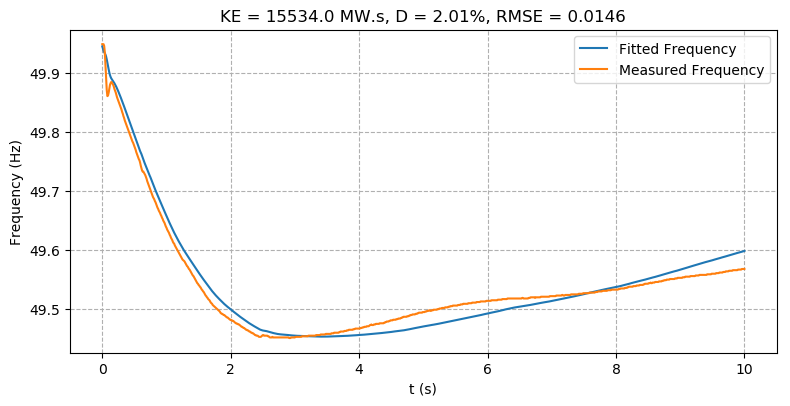}}
    \hfill
  \caption{Case 5: Network fault causing a generator trip}
  \label{fig:case5} 
\end{figure}

The network fault causes a transient component in the frequency trace, which results in overestimates of system inertia with the sliding window method when the window length is small, as shown Fig. \ref{fig:case5a} for window lengths $<$200-ms. This is broadly similar to how the sliding window method performs with the oscillatory frequency trace observed in Case 2. In that previous case, the sliding window method is more accurate beyond a "knee" point window length, where the effects of the oscillations are filtered out. In this case, the knee point occurs at a window length of 210-ms and a system inertia estimate of 14,000 MW.s.

Similarly, the Inoue method will tend to over-fit the frequency trace (and particularly the transient component) at high polynomial order. This can be seen in Fig. \ref{fig:case5b}, where the system inertia estimate is below the known generator inertia at polynomial orders $>$15. The Inoue method does not credibly provide any useful information regarding the correct range for system inertia.

The model fitting results in Fig. \ref{fig:case5c} show reasonably good alignment between the simulated and measured frequency traces ($RMSE=0.0146$) and yields an inertia estimate of 15,534 MW.s. This estimate is broadly consistent with the sliding window method using a window length near the knee point.

\section{Discussion}
The results from the case studies indicate that system inertia estimation using the proposed model-fitting approach is a credible alternative to conventional approaches based on the linearised swing equation. Good alignment between the measured frequency and the simulated frequency implied by the fitted SFR model suggests that the parameters in the SFR model are able to endogenously explain the frequency trajectory. The proposed approach also has the practical advantage of providing a single point estimate without needing to select parameters such as the sliding window length or polynomial order.

Interestingly, the case studies highlight that system inertia estimates using the sliding window and Inoue polynomial fit methods are very sensitive to the choice of these parameters. The computer simulations show that depending on the scenario and choice of window length or polynomial order, system inertia estimates can be either very accurate or have large errors, e.g. $\pm$20\% to 50\%. While there are broad patterns that are observed in the case studies, for example, systematic under-estimates when using the Inoue method with high polynomial orders in frequency traces containing transient or oscillatory components, there are no specific rules or heuristics to guide parameter selection. As the case studies in Section \ref{sec:applications} show, it is difficult to select the most appropriate parameter in practice, particularly as measurement noise present in frequency traces can also affect the inertia estimate.

The proposed model-fitting approach does not suffer from these issues and in the computer simulations, the method yields system inertia estimates within 3\% to 9\% of the true value. 

\section{Conclusion}
This paper proposed a novel approach for simultaneously estimating system inertia and the load relief factor. The proposed approach is based on fitting an SFR model to measured contingency / disturbance data. Case studies demonstrated the applicability and validity of applying the new approach on computer simulations with the IEEE 9-bus and 39-bus systems, and on real events in the SWIS in Western Australia. Furthermore, the proposed method is not sensitive to the selection of algorithm parameters such as the sliding window length or polynomial order.

An avenue for future work is to investigate heuristics for the appropriate selection of sliding window length and polynomial order, potentially using the proposed model-fitting approach to calibrate the range of estimates.

\appendices
\section{Removal of inertial components from generator output traces}
High-speed active power measurements of synchronous generators contain both the inertial response and PFR from the generators. Since inertia is endogenous to models based on the swing equation, the system PFR trace can only be used if the inertial components are removed first (otherwise the effects of inertia are double counted).

\begin{figure}[htp]
\centering
\includegraphics[width=0.7\linewidth]{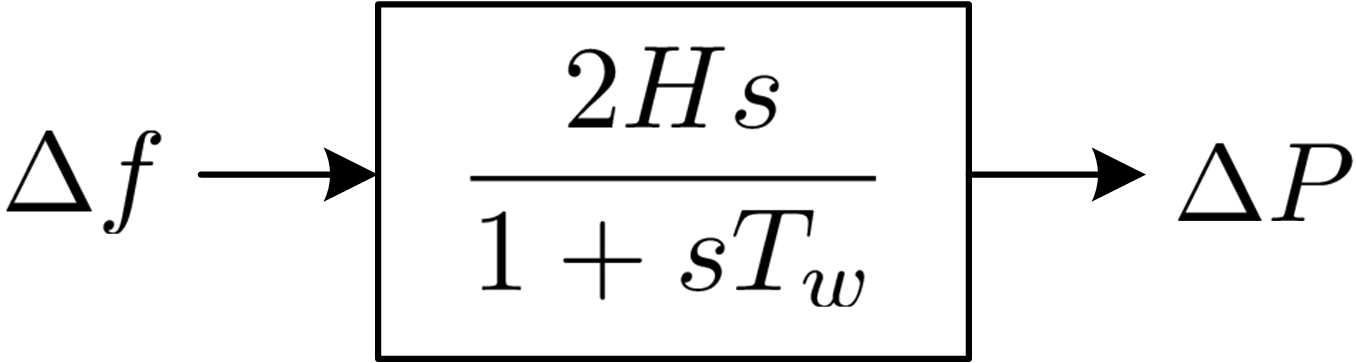}
\caption{Estimating inertial components of PFR with a washout filter}
\label{fig:inertia_removal}
\end{figure}

A simple washout filter is proposed to isolate inertial components from the PFR using the measured frequency as an input and the known inertia constants of the generators contributing to the system PFR as a parameter (see Fig. \ref{fig:inertia_removal}). The inertial components are then removed from the measured system PFR trace.

\begin{figure}[htp]
\centering
\includegraphics[width=1.0\linewidth]{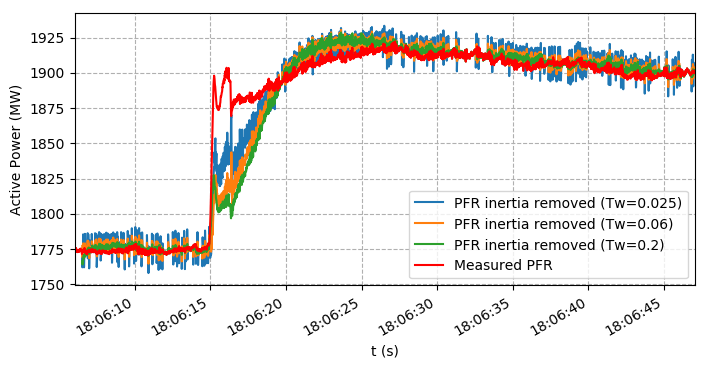}
\caption{Inertia removal with different washout time constants}
\label{fig:inertia_removal_compare}
\end{figure}

The selection of the washout time constant $T_{w}$ is a trade-off between accuracy and noise. Ideally, a very small time constant is applied so that the washout filter more closely approximates a true differentiator. However, small time constants are more sensitive to noise in the frequency measurement (refer to $T_w = 25$-ms in Fig. \ref{fig:inertia_removal_compare}). On the other hand, a large time constant will reduce the accuracy of the estimate despite appearing less noisy (refer to $T_w = 200$-ms in Fig. \ref{fig:inertia_removal_compare}). It was empirically found that a washout time constant of 60 ms to 80 ms yielded an acceptable balance between accuracy and noise (refer to $T_w = 60$-ms in Fig. \ref{fig:inertia_removal_compare}).

\bibliographystyle{IEEEtran}
\bibliography{bib_refs}

\end{document}